\documentclass[amsmath,amssymb,prl,aps,twocolumn,mathbbm,superscriptaddress]{revtex4}
\usepackage{amssymb}
\usepackage{amsmath}
\usepackage{graphicx}
\usepackage{times}
\usepackage{subfigure}

\usepackage[usenames]{color}

\newcommand{\bra}[1]{\langle #1|}
\newcommand{\ket}[1]{|#1\rangle}

\begin{document}
\title{Entanglement of  superconducting qubits via acceleration radiation}
\author{Laura Garc{\'\i}a-{\'A}lvarez}
\address{Department of Physical Chemistry, University of the Basque Country UPV/EHU, Apartado 644, E-48080 Bilbao, Spain}

\author{Simone Felicetti}
\address{Laboratoire Mat\' eriaux et Ph\' enom\`enes Quantiques, Sorbonne Paris Cit\' e, Universit\' e Paris Diderot, CNRS UMR 7162, 75013, Paris, France}

\author{Enrique Rico}
\address{Department of Physical Chemistry, University of the Basque Country UPV/EHU, Apartado 644, E-48080 Bilbao, Spain}
\address{IKERBASQUE, Basque Foundation for Science, Maria Diaz de Haro 3, 48013 Bilbao, Spain}

\author{Enrique Solano}
\address{Department of Physical Chemistry, University of the Basque Country UPV/EHU, Apartado 644, E-48080 Bilbao, Spain}
\address{IKERBASQUE, Basque Foundation for Science, Maria Diaz de Haro 3, 48013 Bilbao, Spain}

\author{Carlos Sab{\'\i}n}
\address{Instituto de F\'isica Fundamental, CSIC, Serrano 113-bis
  28006 Madrid, Spain}

\date{\today}

\begin{abstract}
We show that simulated relativistic motion can generate entanglement between artificial atoms and protect them from spontaneous emission. We consider a pair of superconducting qubits coupled to a resonator mode, where the modulation of the coupling strength can mimic the harmonic motion of the qubits at relativistic speeds, generating acceleration radiation. We find the optimal feasible conditions for generating a stationary entangled state between the qubits when they are initially prepared in their ground state. Furthermore, we analyze the effects of motion on the probability of spontaneous emission in the standard scenarios of single-atom and two-atom superradiance, where one or two excitations are initially present. Finally, we show  that relativistic motion induces sub-radiance and can generate a Zeno-like effect, preserving the excitations from radiative decay. 
\end{abstract}

\maketitle
Circuit Quantum Electrodynamics (cQED)~\cite {Wallraff2004,reviewdevoret,yanamura,reviewnori} offers both a promising architecture for quantum technologies, such as quantum computers~\cite{reviewwilhelm,annealing} and simulators~\cite{cavity,simulatorfermion,digitizedadiabatic}, and a natural arena for the study of quantum field theory and relativistic effects, either in a direct or simulated fashion~\cite{reviewjohansson,rqt,marcos,laura,nonabelian}. For instance, the dynamical Casimir effect (DCE), produced by the modulation of the boundary conditions of the electromagnetic field at relativistic speeds, has been observed in superconducting devices~\cite{moore,casimirwilson,casimirsorin}. Along these lines, it has been shown that DCE radiation possesses several forms of quantum correlations \cite{nonclassical,Benenti2014,discord,ipsteering, Stassi2015} that can be transferred to superconducting qubits \cite{casimirsimone,Rossatto2015}. A related phenomenon is the Unruh effect, where an accelerated detector in vacuum should detect thermal radiation \cite{davieseffect,Milonni94}. Recently, some of us have shown how to mimic the generation of acceleration radiation by means of the modulation of the coupling strength of a superconducting qubit~\cite{relsimone}, a phenomenon resembling the cavity-enhanced Unruh effect~\cite{scully,hureply,scullycomment}. The simulation in a superconducting architecture of both phenomena, DCE and acceleration radiation, relies on the possibility of performing an ultrafast variation of the magnetic flux threading a superconducting quantum interferometric device (SQUID)~\cite{Wallquist06,Johansson10,Andersen2015}.

\begin{figure}[t]
\centering
\includegraphics[angle=0, width=0.4\textwidth]{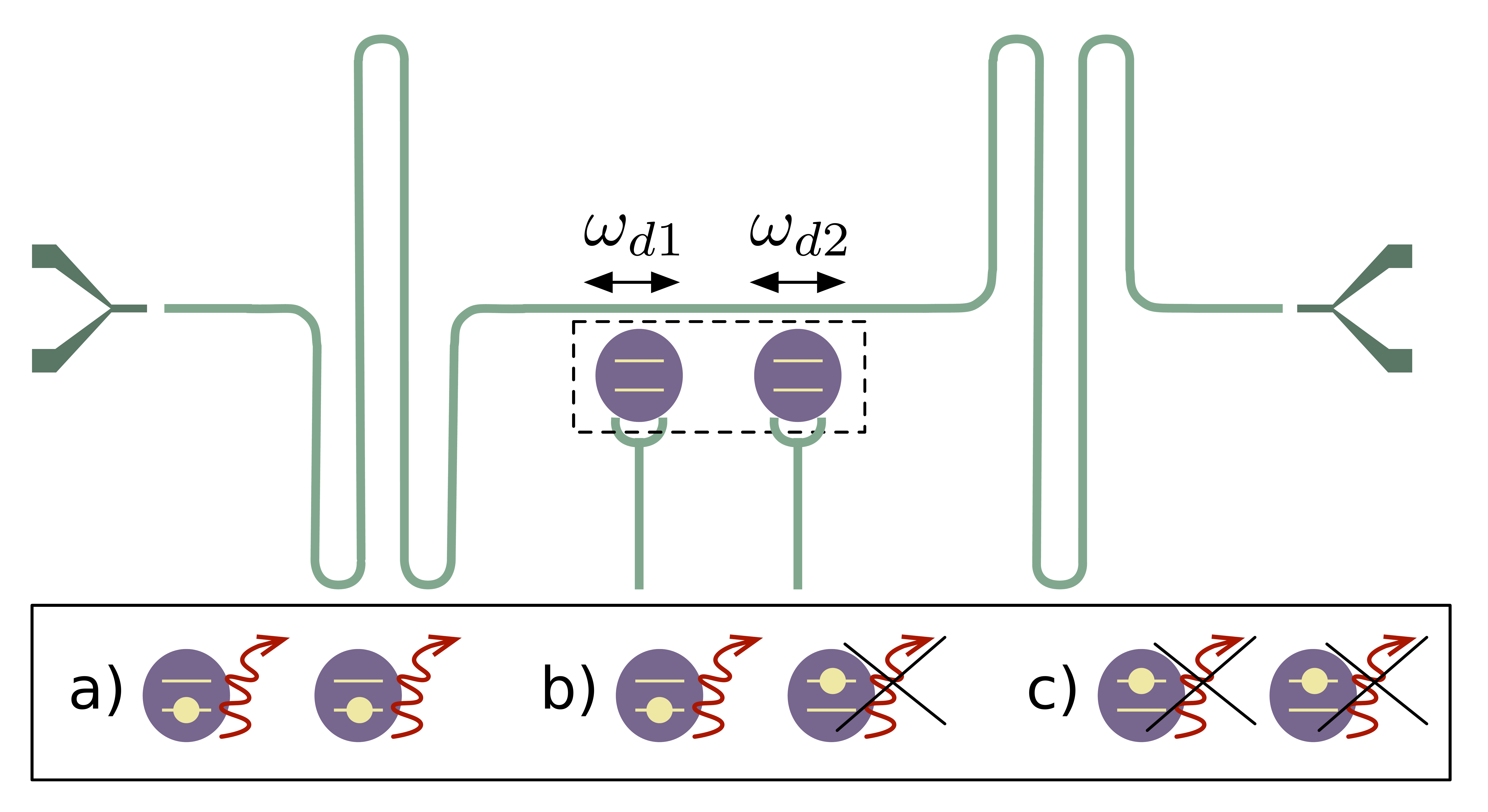}
\caption{Two superconducting qubits strongly coupled to a single resonator mode and driven with frequencies $\omega_{d1}$ and $\omega_{d2}$ simulating harmonic relativistic speeds. The resonator of length $L_c$ is initially in the vacuum while the qubits, which are located in the middle of the cavity $x=0$, are initially a) both in the ground state, b) one in the excited state and the other in the ground state, c) both in the excited state. The red wavy arrows indicate emission or absence of emission of photons from the qubits, showing a subradiant behaviour.}\label{fig1}
\end{figure}

In this paper, we consider a superconducting circuit setup in which two superconducting qubits interact with the same resonator mode and effectively move at relativistic speeds, see Fig.~(\ref{fig1}). The simulation of the relativistic motion of the qubits comes from the modulation of the coupling strength between the qubits and the resonator, which can be interpreted as the qubits movement and activates the counterrotating terms of the quantum Rabi Hamiltonian. We analyse the role of the generated acceleration radiation in several collective properties of the qubits. First, we consider an initial state with no excitations and find the conditions for an efficient generation of stationary entangled states. We find several optimal scenarios for entanglement production: either both qubits move resonantly with the (or half the) natural frequency of the cavity, or one qubit remains static while the other moves at twice the cavity frequency. Second, we analyse the effect of relativistic motion on the spontaneous emission rate when one or two qubit excitations are initially present. Namely, we add the ingredient of relativistic motion to the celebrated Dicke scenario of single-atom and two-atom superradiance~\cite{dique}, which has been recently implemented in a circuit QED architecture~\cite{diquewall}. We will show that the counterrotating dynamics generated by the motion of the qubits tends to suppress the superradiance. Moreover, we find experimental conditions under which the qubit decay is completely frozen, giving rise to a Zeno-like effect induced by the continuous modulation of the coupling strength~\cite{zenosaverio,zenoqubit,zenoreview}. In this second case, the optimal scenario for the appearance of Zeno-like effect correspond to a synchronised motion of the qubits at twice the frequency of the cavity, which generates an anti-Jaynes-Cummings dynamics in both qubits that prevents them from spontaneous emission.

{\textit{Entanglement and acceleration radiation.---}}
The Hamiltonian of the system describes two superconducting qubits of frequency gaps $\omega^q_{\ell}$ coupled to a single resonator mode of frequency $\omega$,
\begin{equation}
\label{hamiltonian}
\mathcal{H}= \hbar \omega a^\dagger a + \sum_{\ell =1}^{2}  \left[  \frac{\hbar\omega^q_{\ell}}{2} \sigma_{\ell}^z + \mathcal{H}_I(x_{q\ell}) \right].
\end{equation}
Here, $\sigma_{\ell}^z$ are Pauli matrices for the qubits, and $a$ ($a^{\dagger}$) is an annihilation (creation) operator for the resonator mode. The interaction Hamiltonian depends on the qubits position as
\begin{equation}
\label{physHI}
\mathcal{H}_I(x_{q\ell}) =  g_{\ell} \cos{\left( k x_{q\ell} \right)} \sigma^x_{\ell} (a^\dagger + a),
\end{equation}
with $g_{\ell}$ the coupling strength and $x_{q\ell}$ the qubit position~\cite{Shanks2013}. In order to simulate the motion of the qubits, which are located in fixed positions, we modulate the coupling with external drivings, such that the interaction Hamiltonian for a qubit reads $\mathcal{H}_I(x_{q\ell}) =  g_{\ell} \cos{\left( f_0 + \Delta f \cos( \omega_{d\ell} t) \right)} \sigma^x_{\ell} (a^\dagger + a)$, and $k x_{q\ell} = f_0 + \Delta f \cos( \omega_{d\ell} t)$. The velocity of the qubits vary harmonically in time, with the maximum value of $\approx \lambda\omega_d\ell$. For $\lambda = 2L_c = 1$cm and $\omega_d\ell = 10$ GHz, we reach values of $\approx 10^8$ m/s $=c/3$.

Initially, we consider the system in the ground state for the qubits and the resonator mode $\ket{g_1\,g_2\,0}$. In order to determine the degree of entanglement between the qubits after a certain interaction time $T$, we compute the concurrence, which up to second order in perturbation theory with respect to $g_{\ell}/\omega$ reads $C=2\,\operatorname{Max}\{|X|-P_e\, ,0\}$~\cite{Sabin2010}. Here, $X$ is the amplitude for photon exchange between the qubits, $X = \langle0|\mathcal{T}(\mathcal{S}^+_1 \mathcal{S}^+_2)|0\rangle $, with $\mathcal{T}$ the time-ordering operator. $P_e$ is the probability of emitting a photon, $P_e = \bra{0}\mathcal{S}^-_1\mathcal{S}^+_1\ket{0}$, with
$
  \mathcal{S}^+_\ell \! = \! - \frac{i\,g_{\ell}}{\hbar}
  \int_0^T
  e^{i\omega^q_{\ell} t'} dt' cos{\left( k x_{q\ell} \right)}  (e^{i\omega t'}a^\dagger + e^{-i\omega t'}a)dt'  = -(\mathcal{S}^{-}_\ell)^\dagger$. In this configuration, if both qubits are at fixed positions, the emission and photon exchange are counterrotating processes, that is, related to the breakdown of the rotating-wave approximation (RWA). These processes will be significant only for ultrastrong couplings or short interaction times. However, the motion of the qubits can excite these counterrotating terms of the Hamiltonian, giving rise to a sizeable emission of photons by a sort of cavity-enhanced Unruh effect~\cite{scully}.
    
\begin{figure}[tbp]
\centering
\includegraphics[angle=0, width=0.45\textwidth]{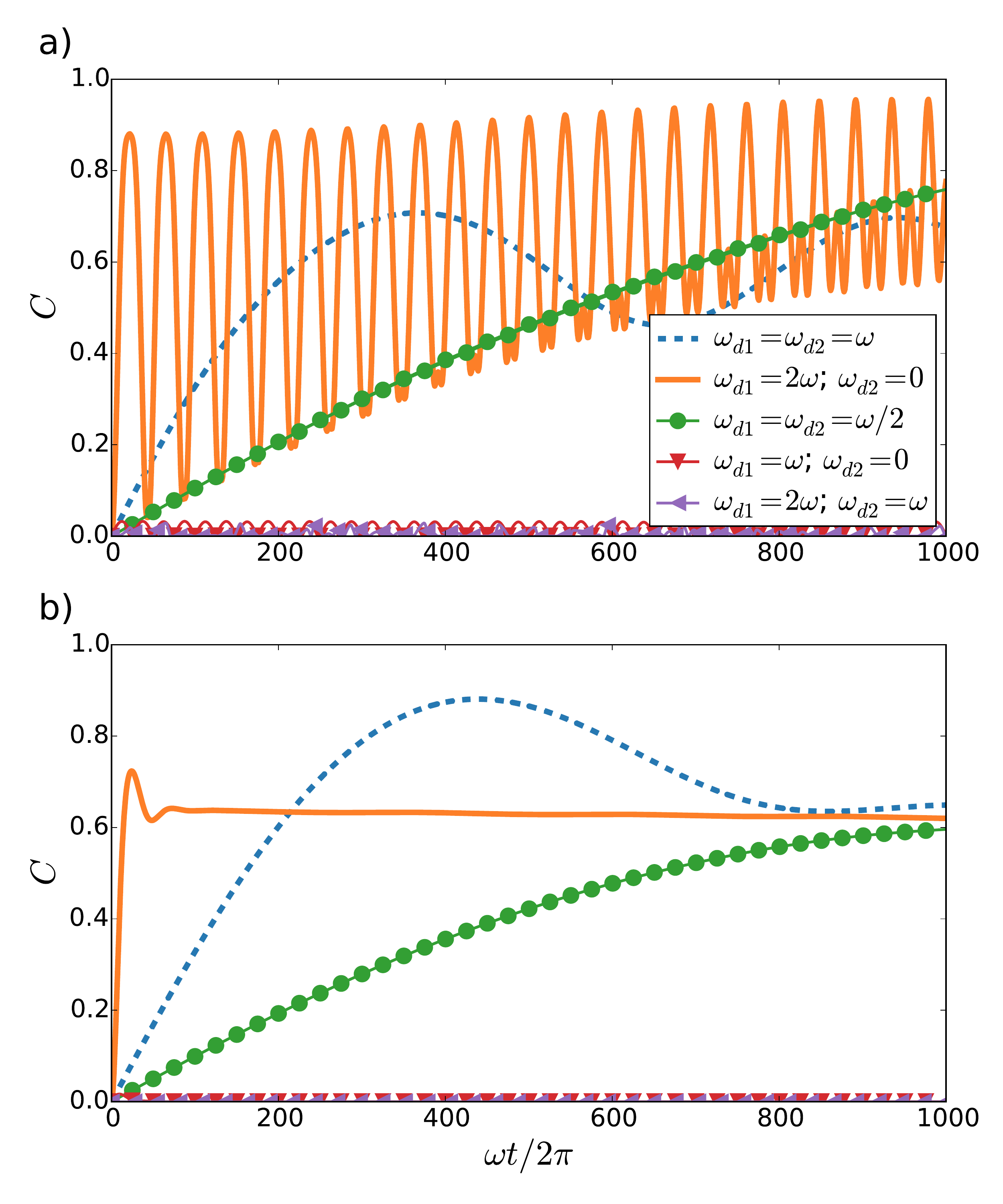}
\caption{Concurrence $C$ of two qubits initially located in the centre of the resonator and oscillating from mirror to mirror with frequencies $\omega_{d1}$ and $\omega_{d2}$, respectively. For coupling constants $g_1=g_2=g=0.02$, initial state $|g_1\,g_2\,0\rangle$, qubit decay parameter $\Gamma = 0.002$, and $T_2/T_1 = 0.67$, we consider two cavity decay rates: a) $\kappa= 0.002$, and b) $\kappa= 0.2$ (bad-cavity limit), in units of $\omega$. These numerical results correspond to a broader parameter range, not limited by the perturbative approximation $gT<1$, which for this case of $g= 0.02$ breaks for $\approx 8\omega t/2\pi$.} \label{fig2}
\end{figure}

We will analyse now under which conditions this phenomenon can be exploited to efficiently generate entanglement between the qubits. We can gain first insights by using analytical techniques. For the sake of simplicity, we assume that $\omega^q_1=\omega^q_2=\omega^q$ and $g_1=g_2=g$. Moreover, for the sake of simplicity, we consider that the harmonic oscillations of the qubits with frequency $\omega_d$ preserve its relative distance, $x_2(t)-x_1(t)=D=\lambda/4$, where $x_1(t)=-\frac{D}{2}(1-\cos{\omega_d\,t})$, and $\lambda$ is the wavelength associated with the cavity frequency $\omega$. Then, we have
\begin{eqnarray}
\label{xpe}
 X &\simeq& g^2\int_0^T dt_2\int_0^{t_2} dt_1 e^{i\,\Delta\,t_2}e^{i(2\omega^q-\Delta)t_1} \nonumber \\&& \cdot \big(J_0(\tfrac{\pi}{2}) -2 J_2 (\tfrac{\pi}{2})\cos{2\omega_d\,t}\big) \nonumber \\
 P_e&\simeq& \frac{g^2}{2}\bigg|\int_0^T dt e^{i(2\omega^q-\Delta)t} \big(J_0(\tfrac{\pi}{4}) \nonumber \\&&  -2 J_2(\tfrac{\pi}{4})\cos{2\omega_d\,t}+2 J_1(\tfrac{\pi}{4})\cos{\omega_d\,t}\big)\bigg|^2,
\end{eqnarray}
where $\Delta$ is the detuning between the qubits and the cavity, $\Delta=\omega^q-\omega$, and $J_n(x)$ are Bessel functions of the first kind. By inspection of Eq.~(\ref{xpe}), we observe that both $X$ and $P_e$ are, in general, oscillating functions. However, under certain resonant conditions, the oscillations are suppressed and these magnitudes grow monotonically in time. For instance, for negligible detuning $\Delta=0$, and frequencies $\omega_d=\omega=\omega^q$, we find that $|X|\simeq\frac{g^2}{2} J_2(\frac{\pi}{2})T^2$ and $P_e\simeq\frac{g^2}{2}J_2(\frac{\pi}{4})^2T^2$, by keeping only the non-oscillating terms. Since $\frac{J_2(\frac{\pi}{2})}{J_2(\frac{\pi}{4})^2}\gg1$, the entanglement grows quadratically in time as
\begin{equation}
C\simeq g^2T^2 \left[J_2(\tfrac{\pi}{2})-J_2(\tfrac{\pi}{4})^2\right].
\end{equation}

Therefore, we predict an entanglement resonance around $\omega_d=\omega$. These analytical results are limited by the perturbative approximation employed, which assumes that $gT <1$. Even in the weak coupling regime, the perturbative approximation would eventually break down. In Fig.~(\ref{fig2}), we plot the results of numerical simulations which generalise our analytical insights. The dynamics is governed by a master equation where we introduce a cavity decay rate $\kappa$, a decay parameter $\Gamma$ accounting for dissipative processes, as well as a decay $\Gamma_{\varphi}$ for the dephasing of the qubits. The energy relaxation time and phase coherence time are denoted with $T_1 = 1/\Gamma$ and $T_2 = 1 / \Gamma_{\phi}$, respectively. We consider realistic parameters, achievable with present technology~\cite{Zhang2016}. This allows us to analyse the long-term dynamics of the system and to consider more general types of motion with $\omega_{d1}\neq\omega_{d2}$, in which the relative distance of the qubits is no longer preserved. Numerical simulations confirm the generation of a high degree of entanglement in the case of $\omega_{d1}=\omega_{d2}=\omega_d=\omega$, as expected for short-time dynamics. In the long-term dynamics, we observe non-trivial entanglement oscillations, where maximum values are achieved at particular times shown in Fig.~(\ref{fig2}). Another optimal scenario for entanglement generation appears when one qubit is effectively moving with frequency $\omega_{d1}=2\omega$, and the other remains static, $\omega_{d2}=0$. Under these circumstances, the first qubit is ruled by an anti-Jaynes Cummings (anti-JC) dynamics which maximises the counterrotating emission of photons~\cite{relsimone}. In this case, the concurrence reaches its maximal value, and the amplitude of initially perfect collapse-revival cycles eventually diminish. Asymptotically, entanglement exhibits small fluctuations around a mean value close to one. Moreover, if we also consider the bad-cavity limit, where $\kappa \gg g \gg \Gamma$, entanglement oscillations are smoothed out, and highly entangled stationary states are reached, see Fig.~(\ref{fig2}b). We extend our analysis of the generation of entanglement between both qubits for the case in which the cavity is out of resonance from both drivings and qubits frequencies. Although we have analytically seen that the concurrence increases when the resonant condition is fulfilled, we expect that considering a detuned cavity will enhance the quantum correlation between the qubits (see Supplementary Information).

\begin{figure}[tbp]
\centering
\includegraphics[angle=0, width=0.45\textwidth]{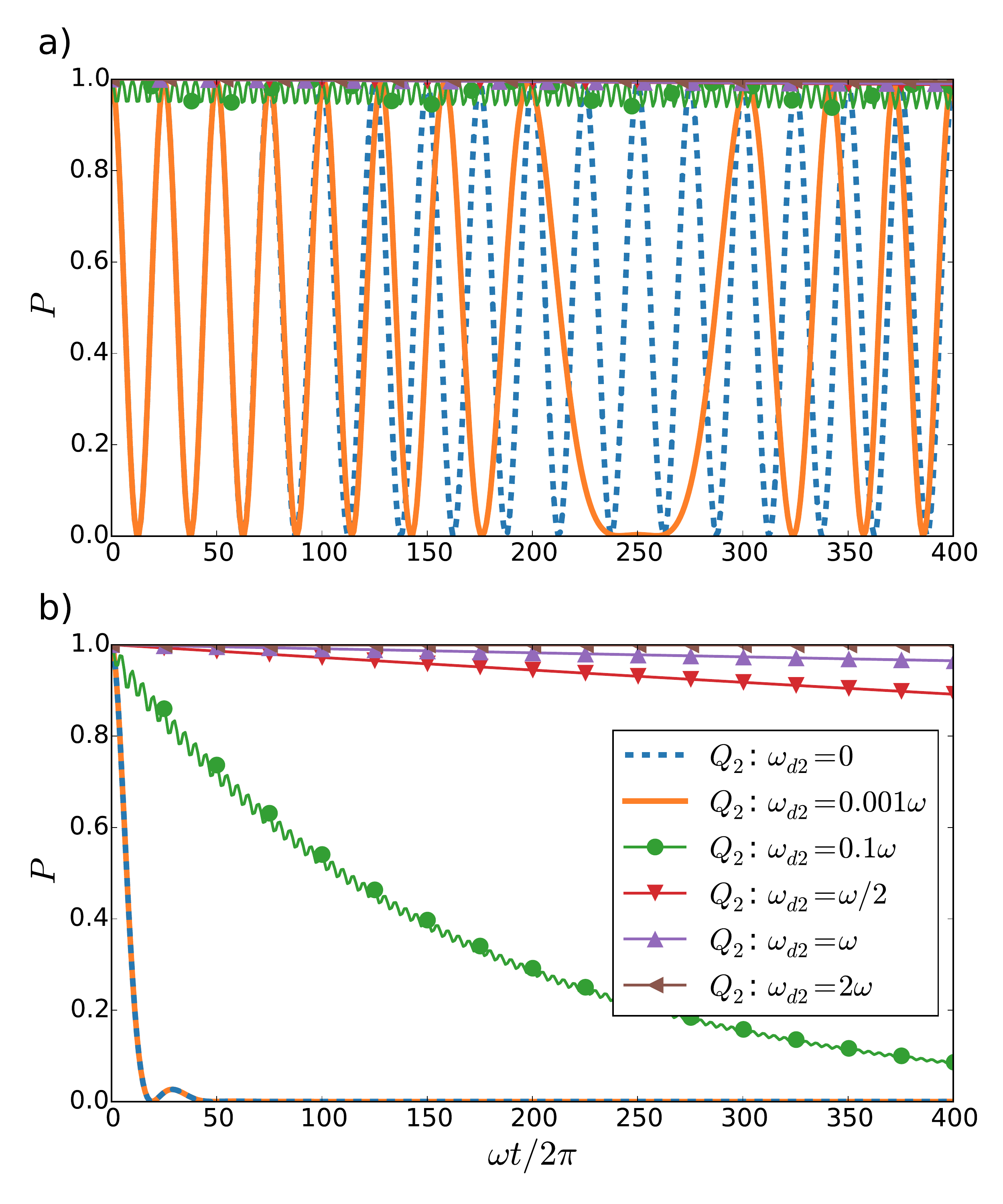}
\caption{Motion effects in single-atom superradiance, observed in the probability $P$ of excitation of the second qubit $Q_2$. We consider the first qubit $Q_1$ decoupled, $g_1=0$, and the second moving with frequency $\omega_{d2}$, for an initial state $|g_1\,e_2\,0\rangle$. We show the behaviour for different velocities of the second qubit, ranging from the static case, $\omega_{d2}=0$, to a velocity of $\omega_{d2}=2\omega$. For a coupling constant $g_2=0.02$, a qubit decay parameter $\Gamma=0.002$ and $T_2/T_1 = 0.67$, we consider two cavity decay rates: a) $\kappa= 0.002$, and b) $\kappa=0.2$ (bad-cavity limit), in units of $\omega$.}  \label{fig3}
\end{figure}

\begin{figure}[tbp]
\centering
\includegraphics[angle=0, width=0.45\textwidth]{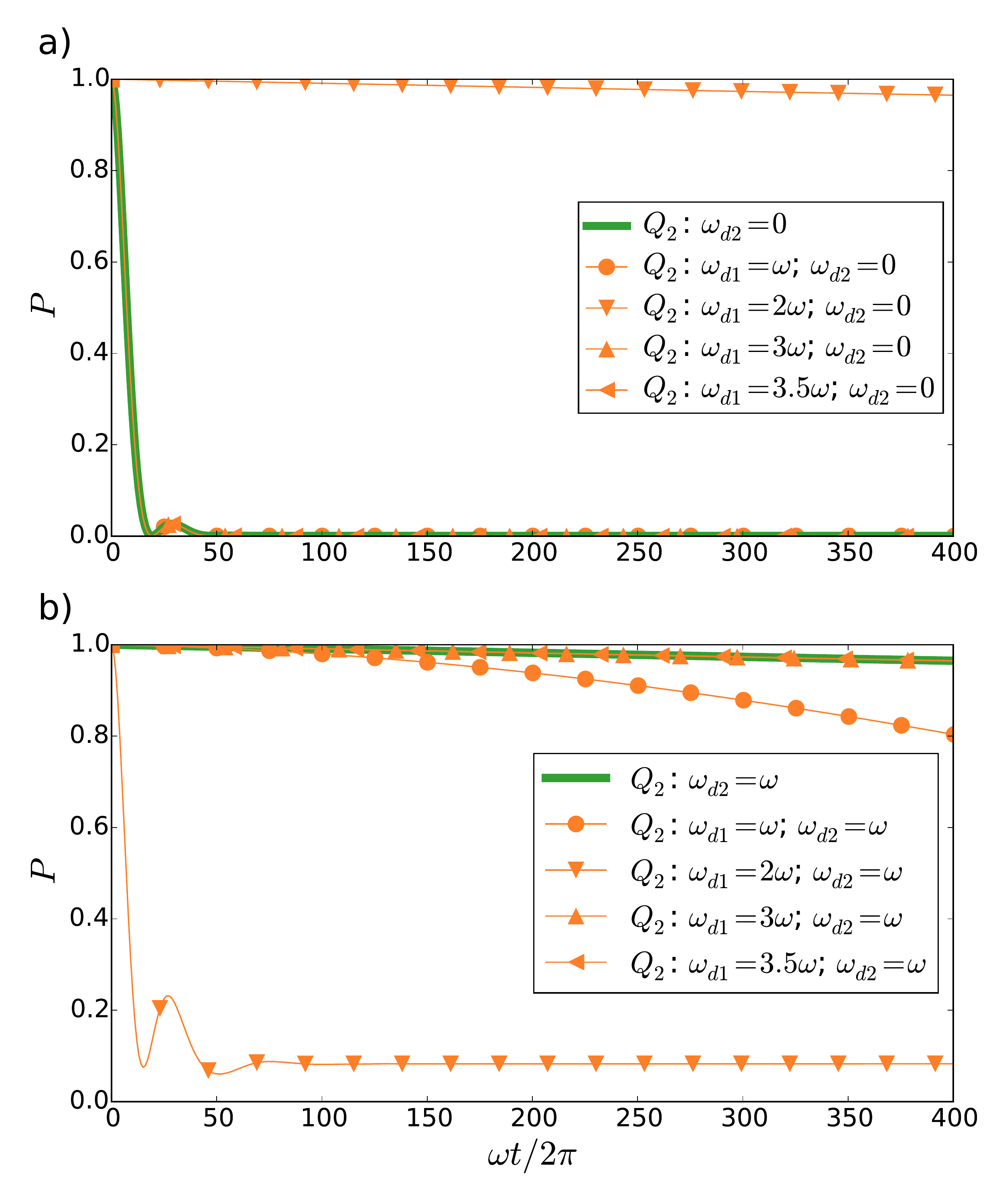}
\caption{Motion effects in single-atom superradiance, observed in the probability $P$ of excitation of the second qubit $Q_2$. We consider the influence of the movement of the first qubit $Q_1$ by analysing different velocities $\omega_{d1}$ and the decoupled case, $g_1=0$. The second qubit is moving with frequency $\omega_{d2}$, for an initial state $|g_1\,e_2\,0\rangle$. We compute for a coupling constant $g_1=g_2=0.02$ in the cases with the first qubit coupled, a qubit decay parameter $\Gamma=0.002$ and $T_2/T_1 = 0.67$, and a cavity decay rate $\kappa=0.2$ (bad-cavity limit), in units of $\omega$. We show the modification in the behaviour in the case of the second qubit a) static $\omega_{d2}= 0$, and b) moving with $\omega_{d2}=\omega$.}  \label{fig4new}
\end{figure}

{\textit{Single-atom superradiance and Zeno-like effect.---}}
In his seminal work, Dicke showed that the decay of atomic emitters is enhanced by the presence of other atomic emitters~\cite{dique}. The simplest case, called single-atom superradiance, consists of a single emitter in an excited state influenced by the proximity of another emitter, even if the latter is in the ground state. This \textit{gedanken experiment} has been recently realised in a circuit QED architecture in the bad-cavity limit \cite{diquewall}. Here, we analyse the effects of relativistic motion in this scenario. To this end, we consider the initial state $\ket{g_1\,e_2\,0}$, and discuss, firstly, the effects of the relativistic motion of the second qubit in its decay, and, secondly, the effects of the presence of the first qubit in the decay of the second one. We observe that the relativistic motion of the second qubit, encoded in $\omega_{d2}$, tends to inhibit its decay leading to a decreased emission rate, known as sub-radiance, see Fig.~\ref{fig3}. Again, we can get some insight on the system dynamics from first-order analytical computations. Since in this case the qubit is initially excited, the probability of emitting a photon is not given by Eq. (\ref{xpe}). In particular, in the resonant case ($\Delta=0$), we have $P_e= g_2^2\bigg|\int_0^T dt \cos{(k\,x_{q2}(t))}\bigg|^2$. Then, considering the trajectories for both qubits such that $x_{q\ell}=L_{c}/2+L_{c}/2\cos{(\omega_{d\ell}\,t)}$ we get $P_e\simeq4g_2^2J_1^2(\pi/2)\sin^2{(\omega_{d2}T)}/\omega_{d2}^2\simeq g_2^2\sin^2{(\omega_{d2}T)/\omega_{d2}^2 }$. This means that for a static qubit -$\omega_{d2}$ close to 0-, the probability of emission grows quadratically in time $P_e\simeq g^2 T^2$, while for frequencies of motion significantly different from 0 the probability oscillates with an amplitude which decreases with $\omega_{d2}$. Therefore, for large enough $\omega_{d2}$ the probability of emission is suppressed, as can be seen in Fig.~(\ref{fig3}). Note that the maximum acceleration of the qubit motion is proportional to $\omega_{d2}^2$. Thus, the larger the acceleration is, the larger the suppression of the probability of emission. Then, the sub-radiance can be seen as another relativistic effect, hitherto unexplored. At first glance, this subradiant dynamics might look surprising, since relativistic accelerated motion is typically associated to the emission of photons. However, note that both phenomena, Unruh effect and subrradiance, are activated by the counterrotating terms of the Hamiltonian, which become dominant for large enough $\omega_{d2}$ associated with relativistic motion. While non-RWA dynamics gives rise to emission of photons when the qubit and cavity start in the ground state, in the present case the initial state $\ket{e_2\,0}$ would be stationary in the presence of only non-RWA terms. The decay dynamics of the second qubit is effectively frozen, that is, we observe a Zeno-like effect generated by the continuous modulation of the qubit-cavity coupling strength, which has an effect similar to a continuous monitoring of the system~\cite{zenosaverio,zenoqubit,zenoreview}.

\begin{figure}[tpb]
\centering
\includegraphics[angle=0, width=0.45\textwidth]{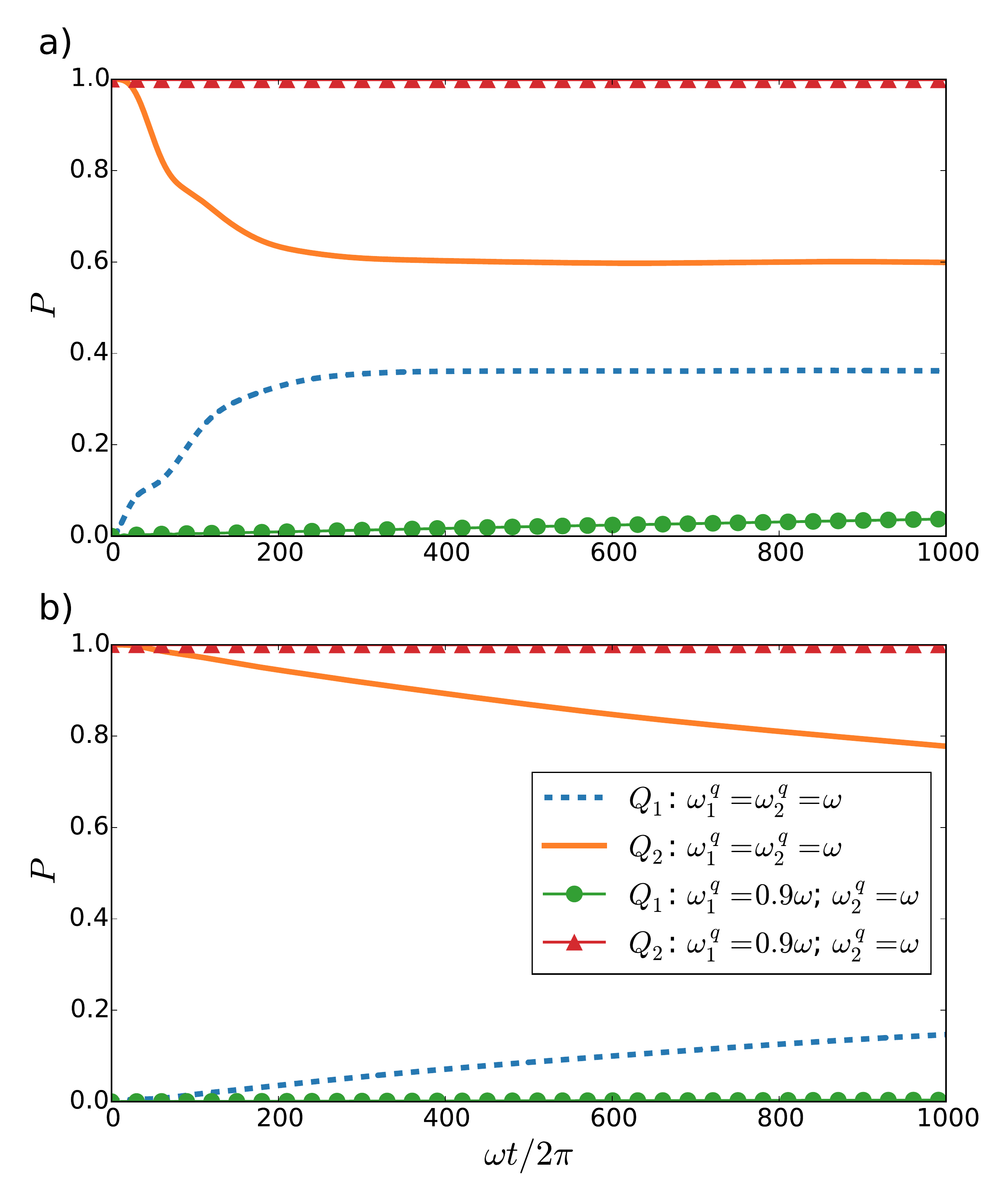}
\caption{Zeno-like effect in the probability $P$ of excitation of two qubits moving with the same frequency $\omega_{d1}=\omega_{d2}=2\omega$ for different amplitudes of oscillation for the first qubit $Q_1$. We consider the initial state $|g_1\,e_2\,0\rangle$, a cavity decay rate $\kappa= 0.1$, coupling constants $g_1=g_2=g=0.01$, and qubit decay parameter $\Gamma=0.001$ and $T_2/T_1 = 0.67$, in units of $\omega$, and that the first qubit is initially placed at $L_c /4$, with $L_c$ the cavity length. We show the excitation probabilities of both qubits, $Q_1$ and $Q_2$, with frequencies $\omega^q_1=\omega^q_2= \omega$, and adding a detuning $\Delta=0.1\omega$ between the first qubit and the cavity, for an amplitude of motion of the first qubit a) $L_c/4$, and b) $L_c/16$.}   \label{fig5}
\end{figure}

Not only the relativistic motion of the second qubit, encoded in $\omega_{d2}$, tends to inhibit its decay, but also the motion of the first qubit, $\omega_{d1}$, has a significant effect on the emission rate for certain values of $\omega_{d1}$. In order to analyse this influence, we compare in Fig.~\ref{fig4new} the probability of excitation of the second qubit for the case in which the first qubit is decoupled, and the case in which it is coupled moving at different relativistic speeds, $\omega_{d1}$. Firstly, we consider the extreme case with the second qubit static, $\omega_{d2}=0$, and we observe that a first qubit relativistic speed $\omega_{d1}=2\omega$ leads to a decreased emission rate, known as sub-radiance, whereas for other combinations of frequencies, the behaviour of the probability of excitation of the second qubit remains unaltered. Secondly, we analyse the case with the second qubit moving with $\omega_{d2}=\omega$, and observe a dramatic change in the decay rate of the second qubit for the same frequency $\omega_{d1}=2\omega$, as in the previous case. We notice that the emission rate of the second qubit is also slightly modified for a relation of frequencies $\omega_{d1}=\omega_{d2}=\omega$ that generates entanglement. However, a further analysis in the relation of the generated entanglement and single-atom superradiance allows us to discard drastic influences of the former in the decay rate (see Supplementary Information). We also interpret the frozen dynamics of the second qubit for $\omega_{d2}=\omega$, and for $\omega_{d2}=\omega$ and $\omega_{d1}=2\omega$, as a Zeno-like effect~\cite{zenosaverio,zenoqubit,zenoreview}. 

We extend our analysis to other scenarios by considering different frequencies and initial conditions. We confirm the Zeno-like effect when we reduce the oscillation amplitude and the excitation probability of the first qubit, which enhances even more the effect on the second qubit (see Fig.~\ref{fig5}). Two-atom superradiance is a collective effect consisting in an enhancement of the decay rate of two emitters with respect to their individual ones~\cite{dique}. We consider the effects of relativistic motion on the decay of two artificial atoms, $\ket{e_1\,e_2\,0}$ and observe a sub-radiance phenomenon, as in the case of single-atom superradiance (see Supplementary Information).

{\textit{Implementation in superconducting circuits.---}}
The model described in Eq.~(\ref{hamiltonian}) can be implemented in a circuit QED architecture~\cite{Blais2004}, using a single-mode transmission line resonator (TLR) interacting with two tunable-coupling superconducting qubits. In order to observe all the phenomenology so far described in a single device, it is required independent tuning of the qubit transition frequencies and of the qubit-cavity coupling strengths. Tunable coupling superconducting qubits~\cite{Gambetta2011, Srinivasan2011,Zhang2016} coupled to a TLR and tuning of effective couplings over nanosecond time-scale~\cite{Bialczak2011} have been proven in circuit QED architectures.

{\textit{Conclusions.---}} 
We have proposed a possible realisation in which simulated relativistic motion generates true entanglement between artificial atoms, while protecting them from spontaneous emission in a Zeno-like effect. A natural extension would be to consider the effects of multi-atom relativistic motion, with a study of superradiant phase transition. Both the ability of generating entanglement and state protection may pave the way for new applications in superconducting quantum technologies. 

\begin{acknowledgments}
This work was supported by a UPV/EHU PhD grant, UPV/EHU EHUA15/17, UPV/EHU UFI 11/55, Spanish MINECO/FEDER FIS2015-69983-P and FIS2015-70856-P, Basque Government grant IT986-16, CAM PRICYT Project QUITEMAD+ S2013/ICE-2801, University Sorbonne Paris Cit\'e EQDOL contract, and Fundaci{\'o}n General CSIC (Programa ComFuturo).
\end{acknowledgments}

\begin{widetext}

\vspace*{0cm}

\section*{SUPPLEMENTARY INFORMATION}

\section{Higher concurrence with off-resonant cavity}
In this section, we extend the analysis of concurrence generation between the pair of moving qubits in the cavity, for the case in which the cavity is detuned from the qubits. We observe that for some specific times in the evolution of the system, the correlation between the qubits is enhanced reaching values close to one. This fact can be explained by considering that the cavity just virtually mediates the interaction between the qubits. As it can be seen in Fig.~\ref{fig1Scon}, this expected feature appears for certain instants of the system time evolution, in a good cavity limit, where the concurrence shows an oscillatory behavior.

\begin{figure}[htbp]
\centering
\includegraphics[angle=0, width=0.9\textwidth]{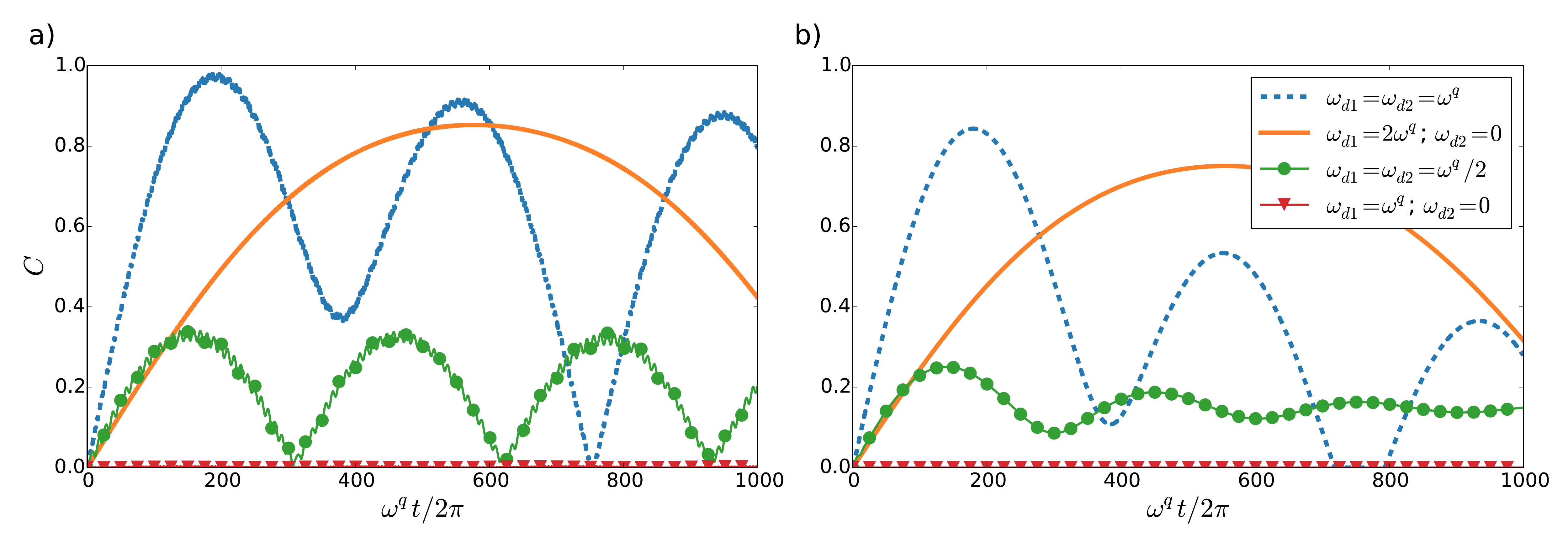}
\caption{Concurrence $C$ of two qubits initially located in the center of the resonator and oscillating with frequencies $\omega_{d1}$ and $\omega_{d2}$, respectively. We consider coupling constants $g_1=g_2=g=0.02$, a qubit decay parameter $\Gamma=0.002$ and $T_2/T_1 = 0.67$, and an off-resonant cavity with $\omega = 0.4$, in units of $\omega^{q}$. For initial state $|g_1\,g_2\,0\rangle$, we compute for two regimes of decoherence, characterized by a cavity decay rate a) $\kappa= 0.002$, and b) $\kappa= 0.2$ (bad-cavity limit), again in units of $\omega^{q}$.} \label{fig1Scon}
\end{figure}

\section{Relation of concurrence and single-atom superradiance}
We have explored numerically the influence of the movement of the first qubit in the emission rate of the second qubit in Fig.~4 of the main text. In order to better interpret the possible relation between the generated entanglement and the influence of the first qubit, we show in Fig.~\ref{fig1S} the concurrence for the relation of frequencies $\omega_{d1}$ and $\omega_{d2}$ analyzed in Fig.~4. We conclude that a combination of frequencies generating entanglement is not relevant for observing significant changes in the probability of the decay of the second qubit, although for $\omega_{d1}=\omega_{d2}=\omega$ it results in a slight modification of the decay rate.

\begin{figure}[hbpt]
\centering
\includegraphics[angle=0, width=0.9\textwidth]{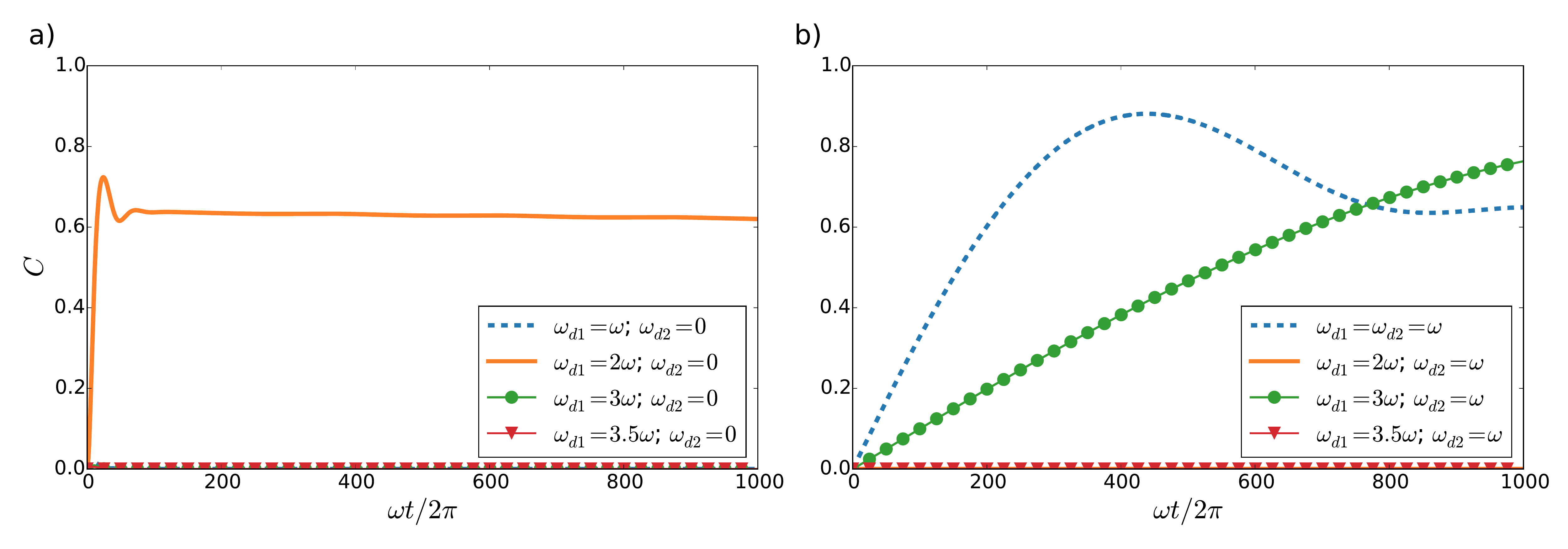}
\caption{Concurrence $C$ of two qubits moving with different velocities $\omega_{d1}$ and $\omega_{d2}$, previously analyzed in the light of single-atom superradiance. We consider a coupling constant $g_1=g_2=0.02$, a qubit decay parameter $\Gamma=0.002$ and $T_2/T_1 = 0.67$, and a cavity decay rate $\kappa=0.2$ (bad-cavity limit), in units of $\omega$. We show results for different velocities $\omega_{d1}$ of the first qubit, when the second qubit is a) static $\omega_{d2}= 0$, and b) moving with $\omega_{d2}=\omega$.}  \label{fig1S}
\end{figure}

\begin{figure}[hbpt]
\centering
\includegraphics[angle=0, width=0.9\textwidth]{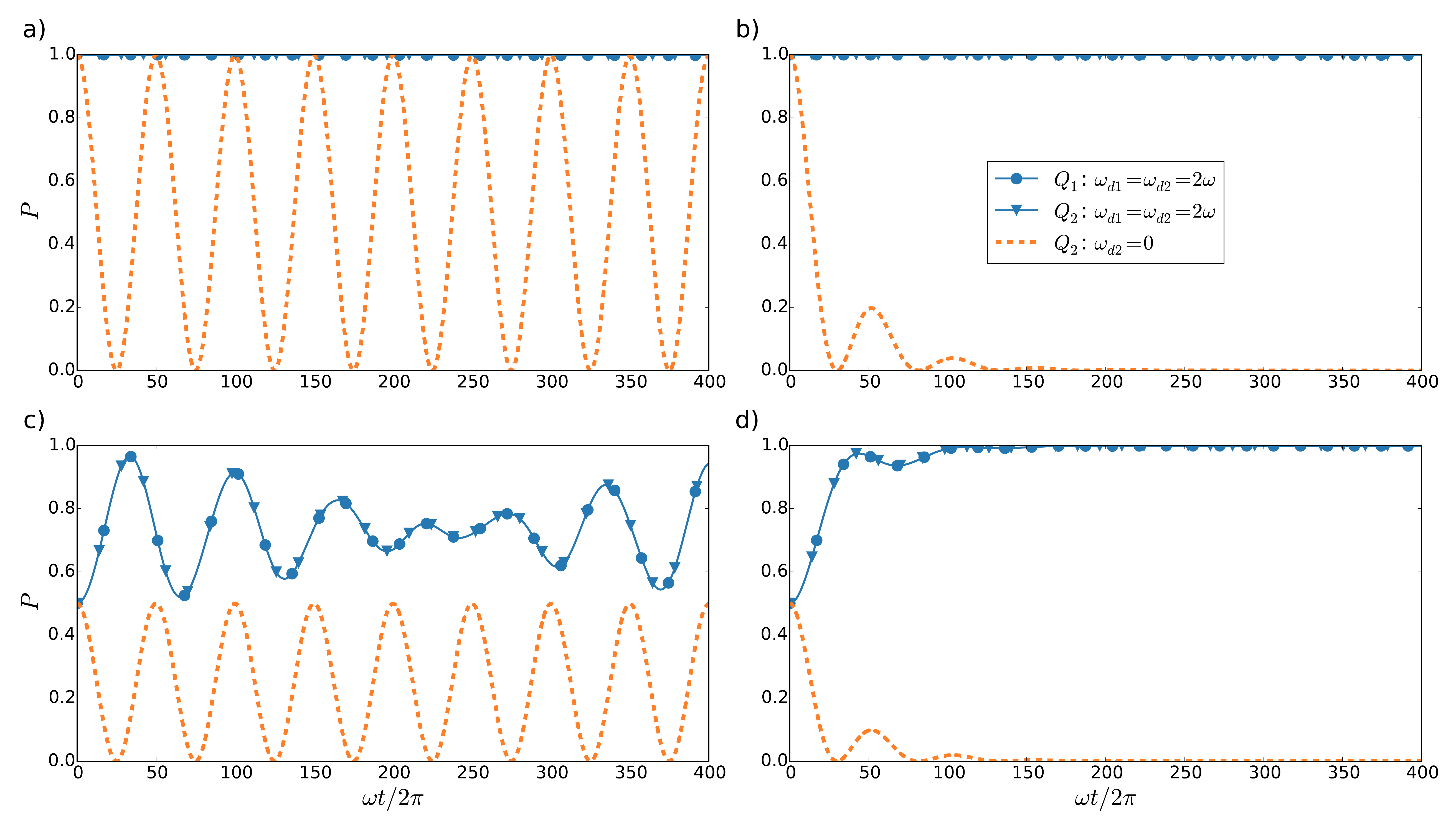}
\caption{Probability $P$ of excitation of two qubits, $Q_1$ and $Q_2$, moving with different frequencies $\omega_{d1}$, $\omega_{d2}$, and a qubit decay parameter $\Gamma=0.001$ and $T_2/T_1 = 0.67$, in units of $\omega$. a), b) Both qubits are initially excited.  c), d) The initial state of the qubits is $\ket{+_1\,+_2}$.  a) , c) $\kappa=0.001$, and  b), d) $\kappa= 0.1$ (bad-cavity limit), in units of $\omega$. All figures show the case of both qubits coupled to the cavity with coupling strengths $g_1=g_2=g=0.01$, $\omega_{d1}=\omega_{d2}=2\omega$, and the case of an uncoupled first qubit $g_1=0$, with a static second qubit $\omega_{d2}=0$ with $g_2=0.01$.} \label{fig2S}
\end{figure}

\section{Two-atom superradiance and collective Zeno-like effect}
We consider the phenomenon of two-atom superradiance in our setup. We compare in Fig.~\ref{fig2S} the decay rate of both qubits moving with relativistic velocities, encoded in $\omega_{d1}$ and $\omega_{d2}$, with respect to their individual decay rates, i.e., with respect to the decay of a single static qubit. We observe that the relativistic motion alters the emission rate and reduces drastically for both qubits the probability of decay, leading to a sub-radiance phenomenon. Both in the good cavity and bad cavity limit, we observe a collective Zeno-like effect, namely the effective freezing of the decay of both qubits. This ties in with the prediction of the expected anti-JC dynamics at this particular driving frequency~\cite{relsimone}. Finally, if we consider a different initial state where both qubits are in a superposition of their ground and excited states, $\ket{+_1\,+_2\,0}$, with $\ket{+}=1/\sqrt{2} (\ket{e}+\ket{g})$, we observe that the qubits are driven fast to their excited states, and the ensuing dynamics follows then the one already observed for an initial $\ket{e_1\,e_2}$.

\section{Implementation details}
With the aim of providing a specific implementation proposal, we focus on a three island superconducting qubit~\cite{Gambetta2011, Srinivasan2011}. The qubit scheme consists in two shunted SQUID loops described by a model composed of two interacting anharmonic oscillators, whose dynamics can be effectively restricted to their two lowest eigenstates. Independent control over the corresponding transition frequencies results in a completely tunable qubit. We denote $\ket{0}_a$ the ground and $\ket{1}_a$ the first excited state of the oscillator $a$, and we call $\omega_a$ the corresponding transition frequency. Same notation will apply for the oscillator $b$. The qubit logical levels are given by the ground $\ket{E_0}$ and first-excited $\ket{E_1}$ states of the collective system, whose structure depends on the ratio between the frequencies $\omega_i$.  When the two anharmonic oscillators are detuned, the lowest collective eigenstates are given by $\ket{E_0} = \ket{0}_a\ket{0}_b$ and $\ket{E_1} = \ket{1}_a\ket{0}_b$, where we set $\omega_a<\omega_b$. In this configuration, the qubit is strongly coupled to the TLR. On the other hand, when the two anharmonic oscillators are nearly degenerate $\omega_a\approx\omega_b$, the first collective excited state is given by $\ket{E_1}= \left( \ket{1}_a\ket{0}_b - \ket{0}_a\ket{1}_b \right) / \sqrt{2}$.  Such state corresponds to an antiparallel configuration of the dipoles of the SQUID loops, hence creating a quadrupolar moment that does not couple with the resonator. Swapping between the two collective energy-configurations, the effective coupling can be continuously tuned in real time, without exciting higher energy levels~\cite{Mezzacapo2014}, which are detuned by at least $1$~GHz from the primary qubit transition~\cite{Gambetta2011}. Using this scheme, tuning of the qubit-cavity coupling  strength has been experimentally achieved in the range from 40 MHz to less than 200 KHz~\cite{Srinivasan2011}.

System initialization is trivial in all the cases considered in this manuscript, as only ground-state cooling and single-qubit gates are required. Fast read-out of the qubit state can be implemented~\cite{Zhang2016} driving a transition to the second-excited collective state $\ket{E_1}\rightarrow\ket{E_2}$. In the resonant case, $\omega_a =\omega_b$, such state is given by $\ket{E_2}= \left( \ket{1}_a\ket{0}_b + \ket{0}_a\ket{1}_b \right) / \sqrt{2}$ and it is strongly coupled to the resonator. The qubit excitation states can be then individually obtained measuring the state-dependent shift on the TLR resonant frequency.

\end{widetext}

\end{document}